%% file: main.tex
\documentclass[9pt, conference]{IEEEtran}
\IEEEoverridecommandlockouts
\usepackage{graphicx} 
\usepackage{booktabs} 
\usepackage{adjustbox} 
\usepackage{cite}
\usepackage{amsmath,amssymb,amsfonts}
\usepackage{algorithmic}
\usepackage{graphicx}
\usepackage{textcomp}
\usepackage{xcolor}
\usepackage{enumitem} 
\usepackage{url} 
\usepackage{tikz}
\usepackage{multirow,cite,bm} 
\usepackage{pifont} 

\def\mv#1{\boldsymbol{{#1}}}

\title{Directional Source Separation for Robust Speech Recognition on Smart Glasses}

\author{Tiantian Feng$^\star$, Ju Lin$^\dagger$, Yiteng Huang$^\dagger$, Weipeng He$^\dagger$, Kaustubh Kalgaonkar$^\dagger$\\ Niko Moritz$^\dagger$, Li Wan$^\dagger$, Xin Lei$^\dagger$, Ming Sun$^\dagger$, Frank Seide$^\dagger$\\ \\

\IEEEauthorblockA{
$^{\star}$University of Southern California,
Los Angeles, USA}
\IEEEauthorblockA{
$^{\dagger}$Meta Platforms, Inc., USA}
}

\begin{document}

\maketitle

\begin{abstract}
Modern smart glasses leverage machine learning to offer real-time transcriptions, considerably enriching human communication experiences. However, such systems frequently encounter challenges related to environmental noises, leading to decreased speech recognition. To improve voice quality, this work investigates directional source separation using the multi-microphone array. We explore multiple beamformers to assist source separation by strengthening the directional properties of speech signals. In addition to relying on predetermined beamformers, we investigate neural beamforming in multi-channel source separation, demonstrating that automatic learning directional characteristics effectively improves separation quality. Furthermore, we investigate the training strategies for ASR when utilizing separated outputs. Our results suggest that jointly training a directional speech separation and ASR model achieves the best overall performance while balancing the wearer and conversation partner's performance.

\end{abstract}

\input{1_intro.tex}

\input{3_ss_modeling.tex}

\input{4_asr_modeling.tex}

\input{5_exp_setups.tex}

\input{6_ss_findings.tex}

\input{7_asr_findings.tex}

\input{8_conclusion.tex}


\bibliographystyle{IEEEbib}
\bibliography{refs}

\end{document}

%% file: 1_intro.tex
\section{Introduction}
\label{sec:intro}

Recent advances in audio sensing \cite{kern2002wearable,feng2018tiles} and augmented reality (AR) \cite{dey2018systematic} have empowered novel applications for smart glasses, enriching the human experience in daily communications by offering robust and efficient speech-understanding systems \cite{somasundaram2023project, he2019streaming}. Specifically, this work explores the microphone array applications on recently introduced \textbf{Project Aria smart glasses}~\cite{somasundaram2023project}, embedded with diverse sensors, including a 7-channel microphone array as shown in Figure~\ref{fig:glasses}. Despite the rich and diverse speech cues the advanced smart glasses capture, these in-the-wild signals are frequently coupled with noises from multiple sources, such as background noises, reverberation, and interfering speakers. Such noises can substantially reduce speech intelligibility, leading to decreased speech recognition. An effective solution to improve voice quality is source separation \cite{wang2018supervised}, which separates relevant speech from ambient sources.

In this paper, we present a comprehensive study of directional source separation on smart glasses. Specifically, we are interested in disambiguating speech by the \textbf{wearer (SELF)}, \textbf{the conversation partner (PARTNER)}, and unrelated bystanders. Here, the wearer is the person wearing the glasses. There are several applications for this technology, such as live captioning and live translation. Unlike prior works \cite{subakan2021attention, luo2020dual} that focus on single-channel setup, our work leverages multi-channel microphones to perform source separation. Multi-channel microphone arrays have advantages over mono-channel setups in providing spatial information to the received speech signals that are beneficial to disambiguating ambient sources. Along with the multi-channel setup, we integrate multiple beamformers as the front-end processor to strengthen the sound sources' directional information. Specifically, this study involves answering the following research questions:

\begin{figure}
    \centering
    \includegraphics[width=0.65\linewidth]{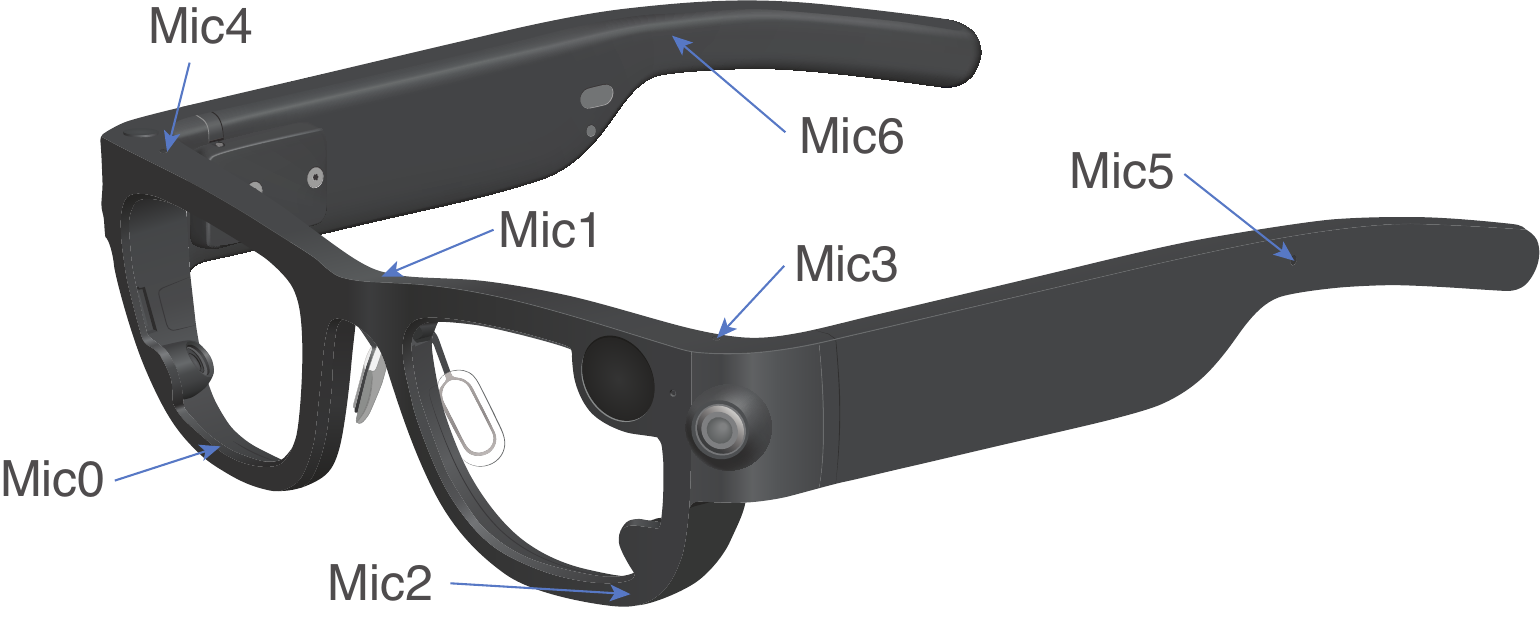}
    \vspace{-2.5mm}
    \caption{7-channel microphone arrays on Project Aria glasses.}
    \label{fig:glasses}
    \vspace{-3mm}
\end{figure}

\noindent \textbf{Is more directional information beneficial for source separation?} In addition to spatial properties embedded in the multi-channel microphone arrays, we propose to utilize the multiple beamformers to enhance directional information from speech signals, enabling the system to implicitly perform speaker disambiguation and noise suppression. Beamforming \cite{benesty2008microphone, capon1969high} is an efficient front-end component aiming to amplify the signal from a specific direction. In a beamformer-based speech pipeline, the beamformer typically provides enhanced speech signals to subsequent systems such as ASR \cite{lin23j_interspeech}.

\begin{figure*}
    \centering
    \includegraphics[width=0.94\linewidth]{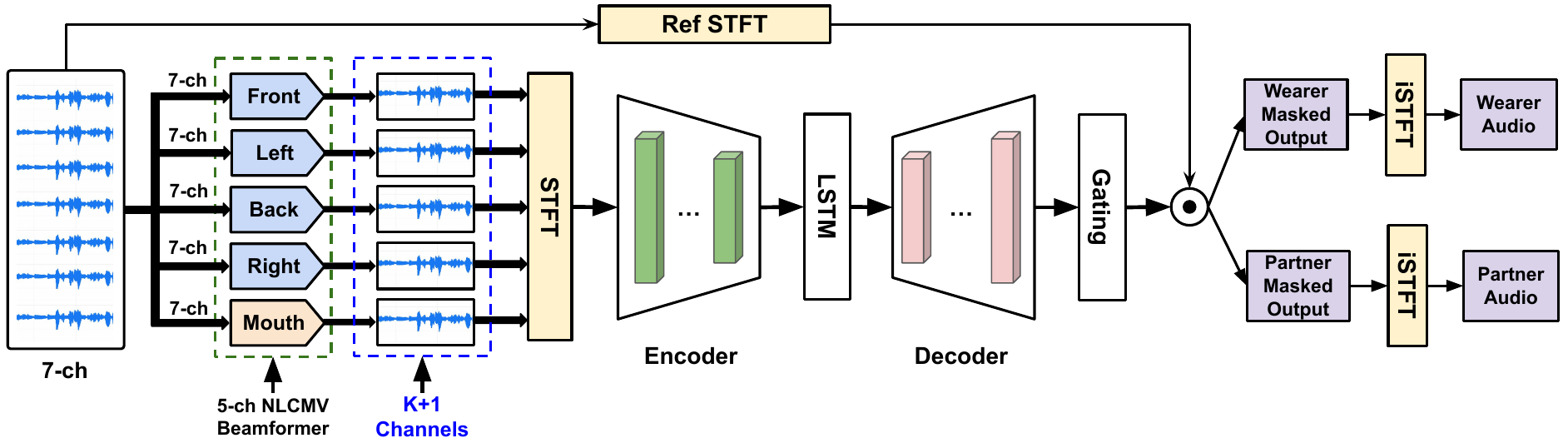}
    \vspace{-2mm}
    \caption{Proposed directional source separation architecture. In the above example, the 7-ch audio input is processed by the beamforming front end, which consists of a 5-ch NLCMV beamformer (steering directions K = 4 and the mouth direction). The source separation back-end takes multi-channel beamformed audio through an encoder-decoder framework.}
    \label{fig:source_separation}
    \vspace{-4mm}
\end{figure*}

\noindent \textbf{Can neural beamforming improve source separation?} Conventional beamformers rely heavily on ideal assumptions about the environments, making them often limited in practice. Instead of drawing unrealistic hypotheses about the environment, neural beamforming \cite{li2016neural, ochiai2017unified, sainath2016factored, he2020spatial, xiao2016deep, braun2019attention} is an emerging technique to learn the beamformer weights from the immense volume of microphone array signals accessible from real-life recordings or by simulation. This motivates us to probe neural beamforming in multi-channel source separation \cite{li2017acoustic,nugraha2016multichannel}.

\noindent \textbf{Can source separation improve speech recognition on smart glasses?}
One primary goal of source separation is to enhance speech recognition. Here, we investigate the impact of source separation on ASR leveraging separation outputs. Increasingly, we explore whether the joint training of the source separation and ASR model would further benefit speech recognition.

\noindent \textbf{In summary, our contributions are summarized as follows: }

\begin{itemize}[leftmargin=*]
    
    \item We investigate the neural beamforming in source separation, discovering that automatic learning directional characteristics open up possibilities for further enhancing voice quality.
    \vspace{-0.1mm}
    
    \item We conduct comprehensive studies quantifying the impact of source separation on ASR. Our results show that source separation benefits the ASR performance for the wearer (1.63\% WER reduction) but decreases speech recognition for the conversational partner. Moreover, combining the separation and beamformed outputs provides competitive ASR performance.

    \item We study joint training of source separation and ASR, demonstrating that joint training achieves the best overall ASR.
\end{itemize}

%% file: 3_ss_modeling.tex
\section{Source Separation Modeling}

Fig~\ref{fig:source_separation} shows the our proposed directional source separation. It consists of front-end multiple beamformers and a source separation neural network. The source separation network receives beamformed outputs and is trained to separate the main speakers, the wearer (SELF), and the partner (PARTNER) in a conversational. The wearer indicates the person wearing the glasses, and the partner is the person who speaks directly to the wearer. The PARTNER
speaker is located at forward-facing angles of -60
to +60 degrees and cross-talk is simulated from other directions. This configuration is labeled V4 in \cite{lin23j_interspeech}.

\vspace{-1mm}
\subsection{Beamforming Front-end}
\vspace{-0.5mm}

In this work, the multiple beamformers preprocess the raw multi-channel audio into $K$ horizontal steering directions around the smart glasses device, plus one in the speaker's mouth direction. Here, we use the predetermined beamformer weights with horizontal steering directions $K=4$ and $K=12$, leading to 5-channel and 13-channel beamformed outputs, respectively. \textbf{In neural beamforming, we treat multiple beamformers as a convolutional layer,} where we load predetermined beamformer weights as the model weights and update the weight using back-propagation. Specifically, we use \texttt{BF} and \texttt{Neural BF} to indicate pre-determined and neural beamformer, respectively.

\vspace{-1mm}
\subsection{Beamforming Design - NLCMV: Non-Linearly Constrained Minimum-Variance
Beamforming}
\vspace{-0.5mm}

As shown in Fig 2, beamforming is one key component of the proposed system. A conventional beamformer algorithm, e.g., Minimum variance distortionless response (MVDR)~\cite{capon1969high}, aims to minimize the estimated beamformer output level while preserving the integrity of the desired signal. However, that approach neglects white noise during optimization and lacks control over null directions. To address these limitations, researchers have recently introduced a novel Non-Linearly Constrained Minimum Variance (NLCMV) beamforming \cite{lin2024agadir}. The NLCMV combines white noise gain and null direction control into its formulation. Here, given the number of
point noise sources $N$, the weight of $n^{th}$ point noise, power
spectral density (PSD) of point noise $\phi_{pp}$, beamformer weights $\mv{h}(j\omega)$ of each steering direction are optimized by minimizing the following: 


\vspace{-2mm}
\begin{eqnarray}
    \scriptsize{
    \mv{h}^H(j\omega)
    \left[\mv{\Phi}_{dd}(j\omega) + \phi_{pp}(\omega)\sum_{n=1}^{N}\alpha_{p,n}\cdot\mv{g}_n(j\omega)\mv{g}_n^{H}(j\omega)\right]\mv{h}(j\omega)
    }
  \label{eq:rec}
\end{eqnarray}
\vspace{-2mm}

which is subject to linear equality where $\mv{h}^H(j\omega)\cdot\mv{g}_n(j\omega)=1$ and a nonlinear inequality constraint that sets the limit on white noise gain $c(\omega) = \mv{h}^{H}(j\omega)\mv{\Psi}(j\omega)\mv{h}(j\omega)<=0$. Moreover, $\mv{\Phi}_{dd}(jw)$ is the covariance matrix of diffuse noise, 
\[ \scriptsize{\mv{\Psi}(j\omega)=\textbf{I} - \mv{g}(j\omega)\mv{g}^{H}(j\omega) \cdot M \left/ \left[\sum_{m=1}^{M}|G_m(j\omega)|^2\right] \right. ,} \] 
where $G_m(j\omega)$ is the channel response from the target speech to the $m$th of $M$ microphones and $\textbf{I}$ is the identity matrix. The details of the beamformer design are described in \cite{lin2024agadir}. In this work, we adopt the NLCMV as our beamforming design.

\vspace{-0.5mm}
\subsection{Source Separation Back-end}
\vspace{-0.5mm}

Our source separation neural network follows an encoder-decoder architecture. From the $K+1$ beamformed channels, we first extract the STFT features. Next, we feed these time-frequency features to the encoder module consisting of multiple convolutional blocks with gated linear units (GLU)~\cite{dauphin2017language} activation function and Dropout layers in between. Subsequently, the encoding output is applied to a 3-layer LSTM, which is then passed to a set of convolutional decoding layers. Then, we send the decoder output to a gating function that returns the STFT masks associated with the wearer and the partner speech from reference audio. In our proposed source separation architecture, we directly apply the first audio channel as the reference audio. Lastly, we compute the masked time-frequency outputs corresponding to the wearer and partner, which are then converted into the wearer and partner speech using the inverse STFT. The optimization objective in source separation modeling combines L1 loss, STFT loss, and Log SI-SDR loss \cite{lin2022speech}.

\begin{figure*}
    \centering
    \includegraphics[width=\linewidth]{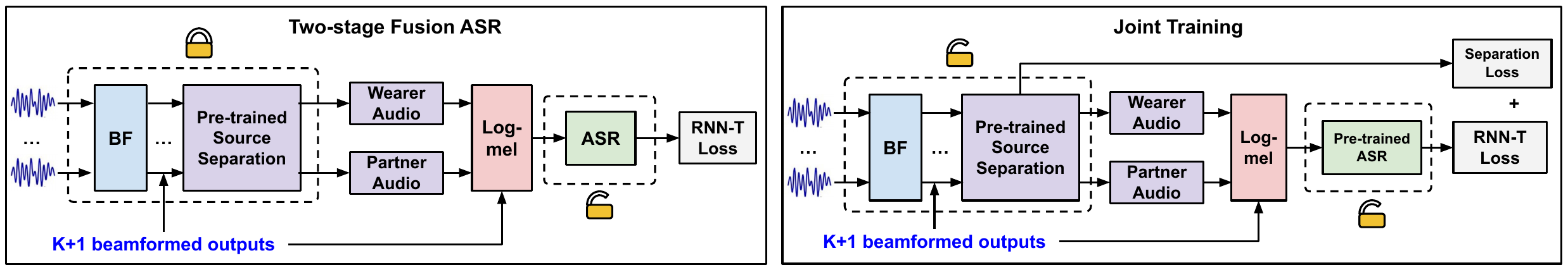}
    \vspace{-4mm}
    \caption{Two stage ASR and joint training ASR. Here, BF is multiple beamformers with K steering directions. In two-stage ASR, only the ASR model is trained, and in joint training, both source separation and ASR models are trained.}
    \label{fig:asr_model}
    \vspace{-3.5mm}
\end{figure*}

%% file: 4_asr_modeling.tex
\section{ASR Modeling}

We further evaluate the source separation modeling through two ASR methods: two-stage ASR and joint ASR modeling.

\subsection{Two-stage ASR Modeling with Source Separation}
The two-stage ASR first extracts the wearer and the partner speech using the pre-trained source separation model. We then compute the log-mel from separate audios, which are subsequently fed into the ASR model. Our ASR network follows the Neural Transducer architecture \cite{mahadeokar2021alignment, moritz2023investigation, sainath2020streaming}, including an encoder, a prediction network, and a joiner network. Our ASR modeling integrates serialized output training (SOT)  \cite{kanda2022streaming, chang2022extended} and uses the alignment-restricted RNN-T loss \cite{mahadeokar2021alignment} as the training objective. In addition to relying solely on separate audio, we study combining the beamformed outputs with the separation output as the ASR input, resulting in $K+3$-channel audio (3=2 separated audio signals + 1 speaker's mouth direction).

\subsection{Joint ASR Modeling with Source Separation}
We investigate the joint training of ASR and source separation in addition to two-stage ASR training. Instead of training both models from scratch, we load the pre-trained ASR weights from two-stage ASR training and pre-trained source separation weights. We combine the ASR training objective with source separation loss to optimize both models in joint training.

%% file: 5_exp_setups.tex
\vspace{-1.5mm}
\section{Experimental Setups}

\subsection{Dataset Details}

We conduct experiments using the open-source Librispeech corpus, which consists of 960 hours of speech from audiobooks in the LibriVox project \cite{panayotov2015librispeech}. To simulate the training data, we generate 100,000 multichannel room impulse responses (RIRs) for rooms with sizes ranging from [5, 5, 2] to [10, 10, 6] meters. We apply the geometry of Aria glasses to simulate multi-channel data. Aria has 7 microphones. We generate the multi-channel signals using image-source methods (ISM) \cite{lehmann2008prediction}. The data simulation framework is adapted from ~\cite{lin23j_interspeech,lin2024agadir}. To better understand the impact of cross-talk and background noises on speech recognition, we generate several test scenarios varying the number of bystanders and SNR range, each containing 3367 utterances from Librispeech.

Furthermore, we add noise from the public noise set~\cite{reddy2020interspeech} to the clean audio segments in both the training and test sets. The SNRs of the mixed audio range from -8 dB to 40 dB relative to the combined audio of the wearer and partner, with an incremental level of 1 dB. Increasingly, we select overlap ratios between the bystanders and the primary speakers, ranging from 5\% to 50\%. With an overlap ratio of 0\%, there is no overlap between bystanders and main speakers (wearer and partner).

\begin{figure}
    \centering
    \includegraphics[width=0.95\linewidth]{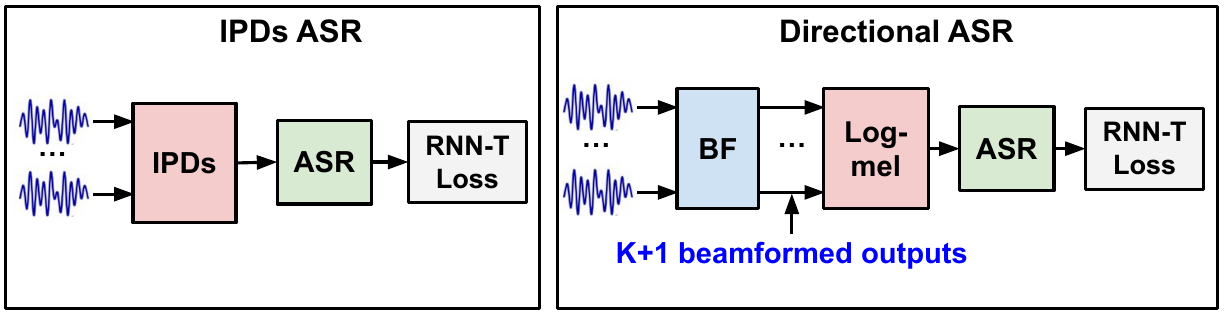}
    \vspace{-3.mm}
    \caption{ASR baselines. BF is multiple beamformers.}
    \label{fig:asr_baseline}
    \vspace{-3.5mm}
\end{figure}

\subsection{Directional Source Separation Training}

Our source separation training uses Librispeech data. We perform different source separation training with the same architecture except for a different input dimension. We extracted 257-dimensional complex SFTF for each beamformer direction or raw microphone channel. Input features from multiple directions or channels are concatenated. We use an Adam optimizer with a tri-stage learning-rate scheduler. We trained the source separation models for 60 epochs, with a learning rate of 4e-4, a warmup of 10k iterations, and forced annealing after 10 epochs.

\subsection{ASR Training}
\vspace{-0.5mm}

We perform the ASR training in baseline and two-stage systems for 120 epochs. Like source separation training, ASR training uses an Adam optimizer, a base learning rate of 0.001, and a warmup of 10,000 iterations. On the other hand, we perform 30 epochs of joint training with the pre-trained ASR model and pre-trained source separation model on Librispeech data. We choose a learning rate of 1e-4 and use the equal weights of 1 in combining ASR loss and the source separation loss. We further compare ASR modeling relying on source separation with the following baseline ASR methods, as shown in Figure~\ref{fig:asr_baseline}:

\noindent \textbf{Directional ASR:} We apply the directional ASR reported in \cite{lin23j_interspeech} as a baseline. Like the beamformer source separation, the directional ASR processes the multi-channel audio into $K+1$ beamformed channels, which are then fed into the ASR model. 

\noindent \textbf{Interchannel phase differences (IPDs): } In addition to directional ASR, we design the baseline model using IPD features as the ASR input. IPDs capture the variations in phase between different audio channels, providing spatial properties of sound sources. We use the IPDs ASR system implemented in \cite{lin23j_interspeech}.


\begin{table}[t]
    \caption{Source separation performance comparisons, where the evaluation set includes only one bystander.}
    \vspace{-1mm}
    \footnotesize
    \centering
    \begin{tabular*}{\linewidth}{lcccc}
        \toprule
        &
        \multicolumn{2}{c}{\textbf{PESQ}} &
        \multicolumn{2}{c}{\textbf{SI-SDR (dB)}} \\

        &
        \multicolumn{1}{c}{\textbf{Wearer}} &
        \multicolumn{1}{c}{\textbf{Partner}} &
        \multicolumn{1}{c}{\textbf{Wearer}} &
        \multicolumn{1}{c}{\textbf{Partner}} \\

        \cmidrule(lr){1-1} \cmidrule(lr){2-3} \cmidrule(lr){4-5}
        \textbf{Without BF - Baseline} & $2.89$  & $1.80$ & $18.17$ & $8.50$ \\

        \midrule
        \texttt{\textbf{BF-5}} & $2.88$  & $1.82$ & $18.09$ & $8.55$ \\
        \texttt{\textbf{BF-13}} & $2.95$  & $1.86$ & $18.33$ & $8.83$ \\

        \midrule
        \texttt{\textbf{Neural BF-13(Ours)}} & $\mathbf{3.11}\uparrow$  & $\mathbf{1.89}\uparrow$ & $\mathbf{20.44}\uparrow$ & $\mathbf{9.51}\uparrow$ \\

        \bottomrule
    \end{tabular*}
\vspace{-4.5mm}
\label{tab:source_separation}
\end{table}

\begin{figure}
    \centering
    \begin{tikzpicture}

        \node[draw=none,fill=none] at (0,0){\includegraphics[width=0.8\linewidth]{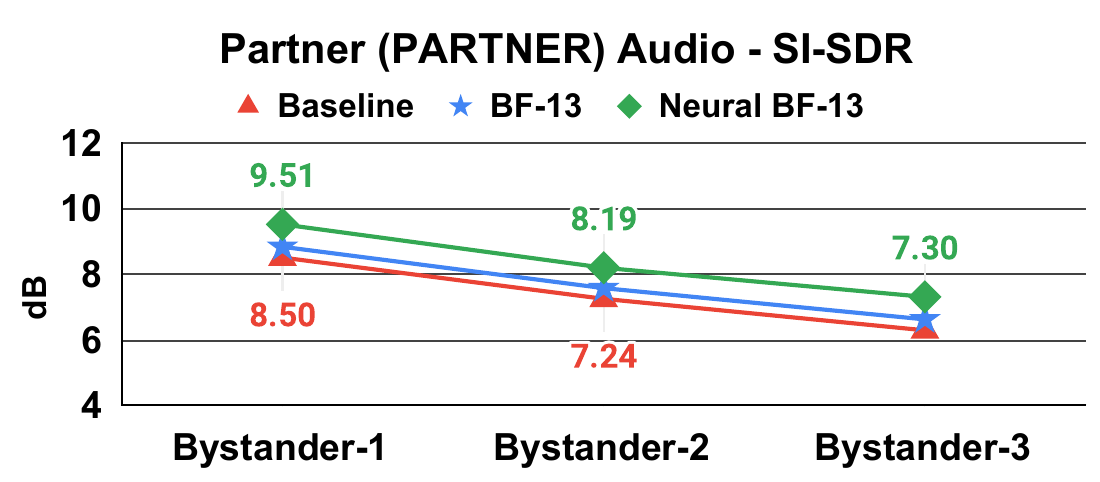}};
        
        \node[draw=none,fill=none] at (0, 2.9){\includegraphics[width=0.8\linewidth]{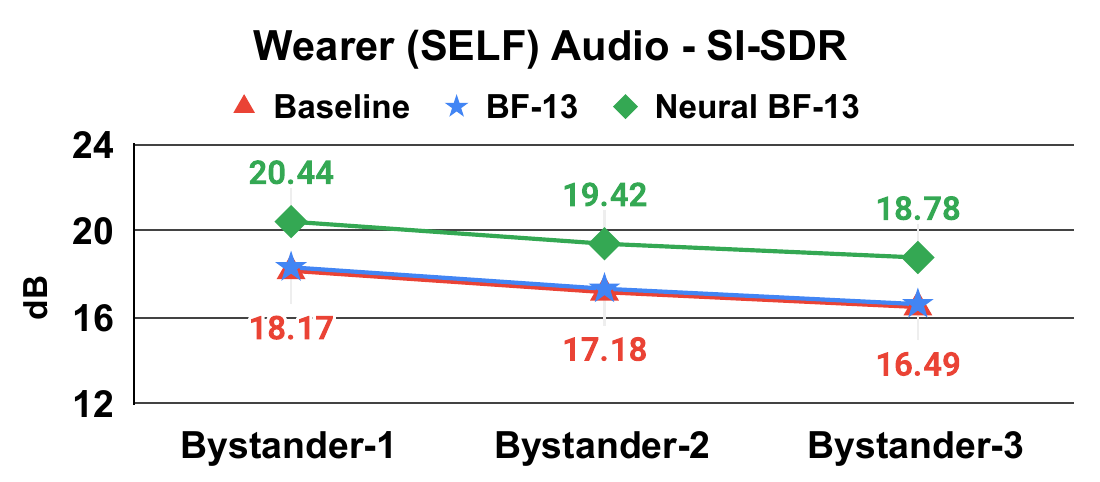}};

    \end{tikzpicture}
    \vspace{-3mm}
    \caption{Source separation varying number of bystanders.}
    \label{fig:separation_distractors}
    \vspace{-3mm}
\end{figure}

%% file: 6_ss_findings.tex
\vspace{-1mm}
\section{Source Separation Findings}
\vspace{-0.15mm}
\subsection{Directional Information Benefits Source Separation}
We first compare the source separation with and without directional information in Table~\ref{tab:source_separation}, quantified using the perceptual evaluation of speech quality (PESQ) \cite{rix2001perceptual} and Scale-Invariant Signal-to-Distortion Ratio (SI-SDR) \cite{le2019sdr}. The baseline system employs a similar encoder and decoder architecture as depicted in Fig.\ref{fig:source_separation}, but it utilizes the complex short-time Fourier transform (STFT) features of 7-channel raw microphone signals without beamforming. We refer to the directional source separation as \texttt{BF-5} and \texttt{BF-13}, where 5 and 13 indicate the number of beamformed channels. The results show that increasing beamforming directions improves the source separation, and \texttt{BF-13} achieves approximately 0.25 dB performance gain over \texttt{BF-5} in both wearer and partner output. Moreover, \texttt{BF-13} with directional information yields higher voice quality than the baseline multi-channel source separation in both PESQ and SI-SDR.

\begin{figure}
    \centering
    \includegraphics[width=\linewidth]{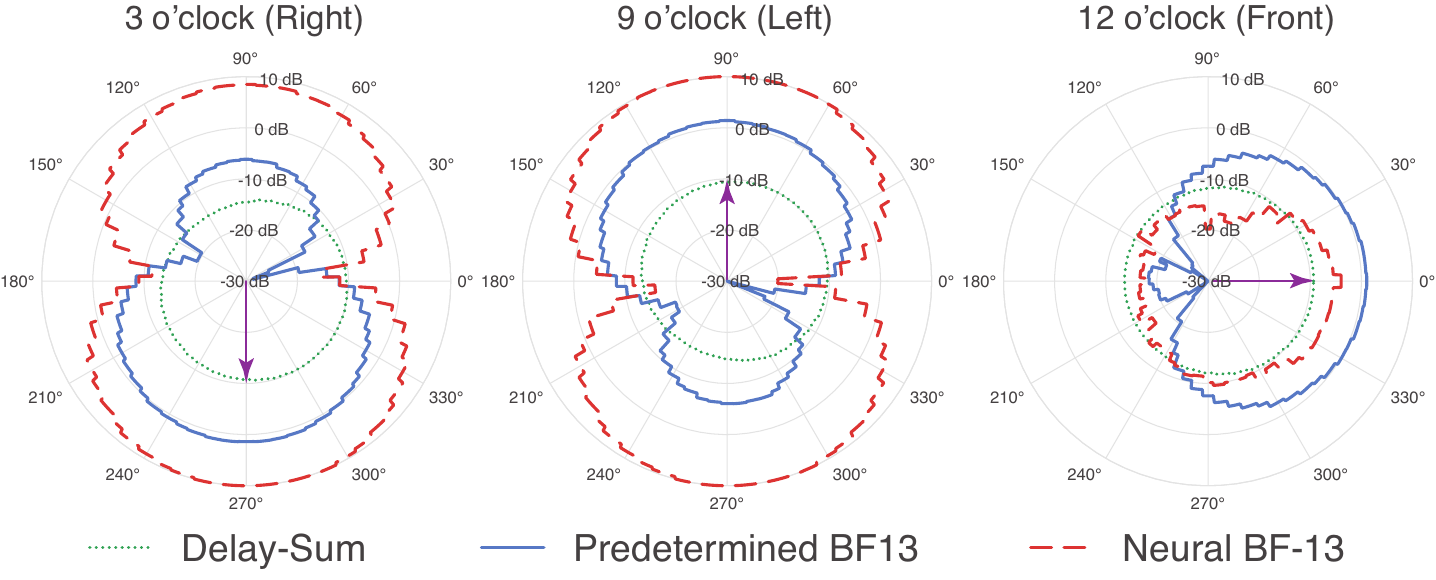}
    \vspace{-4.5mm}
    \caption{Beam patterns at $f = 250$ Hz}
    \label{fig:beam_pattern}
    \vspace{-1.5mm}
\end{figure}

\subsection{Improve Source Separation by Neural Beamforming}

Next, we study the source separation using neural beamforming, as shown in Table~\ref{tab:source_separation}. We find that
\texttt{neural BF-13} substantially improves separation quality, leading to a 2.27 dB and a 1.01 dB SI-SDI increase in the wearer and partner signals, respectively. We also observe an increase in PESQ for wearer and partner separation outputs. In addition to evaluating the condition with one bystander, we compare the voice quality varying (2 and 3) bystanders nearby. The results in Fig.~\ref{fig:separation_distractors} show that \texttt{neural BF-13} is more robust against increased bystanders than multi-channel and non-neural beamformer methods.

\vspace{-1.5mm}
\subsection{Interpreting Neural Beamforming}

Neural beamforming offers encouraging source separation performance, as shown in Table~\ref{tab:source_separation}, but it is unclear why it improves source separation. To unfold what neural beamforming learns, we perform a detailed beamformer analysis as shown in Fig~\ref{fig:beam_pattern}. The plot depicts the beam patterns of the beamformers in 3 distinct directions. We find that neural beamformers have substantial gains (over 10 dB) in the lateral directions (left and right), potentially leading to improved overall source separation.

\begin{table}[t]
    \caption{ASR performance comparisons on LibriSpeech data, where the evaluation set includes only one bystander.}
    \vspace{-2mm}
    \footnotesize
    \centering
    \begin{tabular*}{\linewidth}{lcccc}
        \toprule

        \multicolumn{1}{c}{\textbf{ASR}} &
        \multicolumn{1}{c}{\textbf{Source}} &
        \multicolumn{3}{c}{\textbf{WER}} \\

        \multicolumn{1}{c}{\textbf{Training}} &
        \multicolumn{1}{c}{\textbf{Separation}} &
        \multicolumn{1}{c}{\textbf{Overall}} &
        \multicolumn{1}{c}{\textbf{Wearer}} &
        \multicolumn{1}{c}{\textbf{Partner}} \\

        \cmidrule(lr){1-1} \cmidrule(lr){2-2} \cmidrule(lr){3-5}
        \textbf{IPDs} & N.A.  & $15.12$ & $7.99$ & $22.33$ \\
        \textbf{Dir. ASR BF-13} & N.A.  & $14.14$ & $8.28$ & $20.12$ \\

        \midrule
        \textbf{Two-stage} & Without BF  & $17.28$ & $6.79$ & $27.69$ \\
        \textbf{Two-stage} & \texttt{BF-13}  & $17.06$ & $7.13$ & $27.07$ \\
        \textbf{Two-stage} & \texttt{Neural BF-13}  & $16.04$ & $\mathbf{6.51}$ & $25.46$ \\
        \textbf{Two-stage Fusion} & \texttt{Neural BF-13}  & $13.70$ & $6.65$ & $20.66$ \\

        \midrule
        \textbf{Joint Train Fusion} & \texttt{Neural BF-13}  & $\mathbf{13.25}$ & ${8.06}$ & $\mathbf{18.89}$ \\

        \bottomrule
    \end{tabular*}
    \vspace{-3.5mm}
\label{tab:asr}
\end{table}

%% file: 7_asr_findings.tex
\vspace{-0.5mm}
\section{ASR Findings}
\subsection{Source Separation Benefits ASR on Smart Glasses}

This paragraph compares the ASR performance between baselines and two-stage ASR systems, as demonstrated in Tab.~\ref{tab:asr}. Here, two-stage fusion refers to the two-stage ASR that combines beamformed outputs with the separation outputs, and the two-stage represents the ASR model using only separation outputs. The results show that two-stage ASR outperforms IPDs and directional ASR for wearer speech, indicating improved quality of the wearer speech from source separation. We also observe that the source separation with neural beamforming (\texttt{Neural BF}) yields lower WER than baseline source separation (without BF). However, partner speech suffers a substantial WER increase among all two-stage systems. This result implies far-field speech separation remains a challenging task, and the relatively lower separation quality for partner speech causes performance decreases in the ASR. We perform the two-stage fusion ASR modeling (with \texttt{neural BF-13}), which combines raw multi-channel audio with separation output, to resolve this quality issue from source separation, resulting in overall and wearer ASR performance improvements (WER: 13.70\%) compared to the directional ASR system (WER: 14.14\%).







\begin{figure}
    \centering
    \begin{tikzpicture}

        \node[draw=none,fill=none] at (0,3.3){\includegraphics[width=0.9\linewidth]{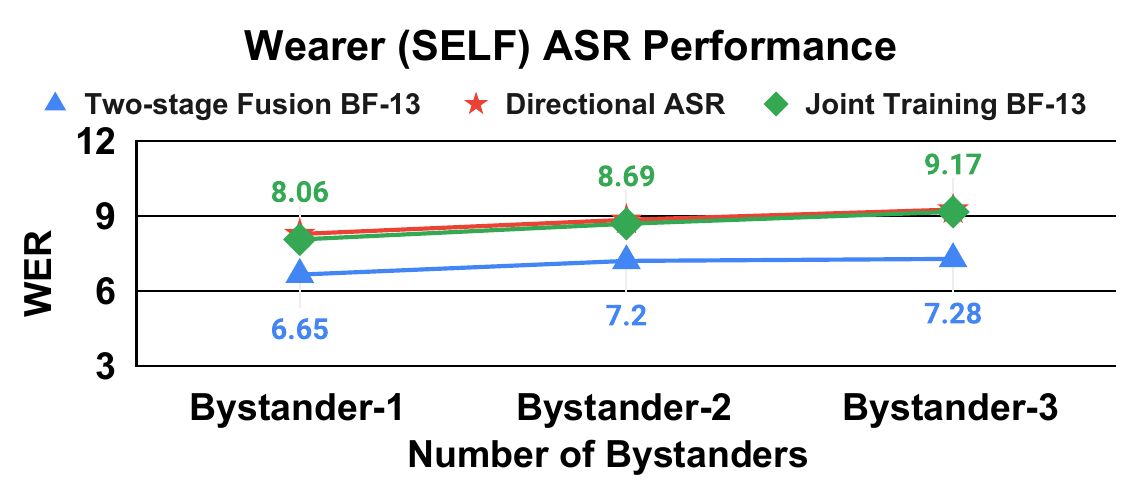}};

        \node[draw=none,fill=none] at (0,0){\includegraphics[width=0.9\linewidth]{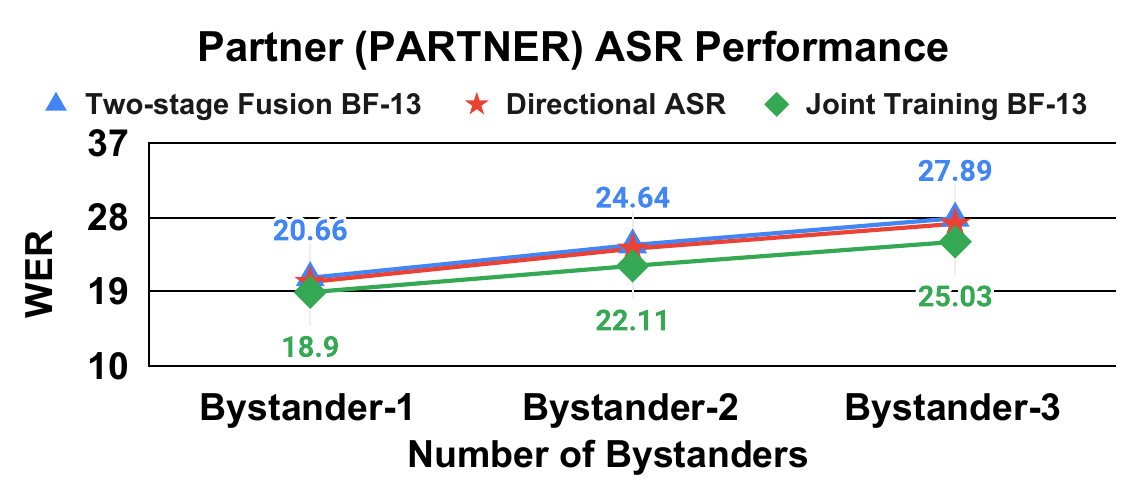}};

    \end{tikzpicture}
    \vspace{-3mm}
    \caption{ASR performance varying number of bystanders.}
    \label{fig:asr_distractors}
    \vspace{-3mm}
\end{figure}
\vspace{-2mm}
\subsection{Improve ASR Performance by Joint Training}

Table~\ref{tab:asr} reveals that joint training with the best ASR system from the two-stage fusion training (with \texttt{Neural BF-13}) would improve ASR performance, reducing overall WER to 13.25\% compared to 14.14\% using directional ASR.
Increasingly, we identify that joint training yields an increased WER for the wearer but reduces WER for partner speech compared to two-stage ASR. Unlike source separation modeling that only yields ASR improvements for wearer speech, joint training strikes a delicate balance in improving speech recognition in all main speakers. We further compare the ASR performance using joint training to other ASR models with varying bystanders, as shown in Figure~\ref{fig:asr_distractors}. The comparisons indicate that joint training also yields more robust partner ASR performance with increasing bystanders, creating a larger relative WER difference (WER difference increases from 1.76\% to 2.86\%) with 3 bystanders.

%% file: 8_conclusion.tex
\section{Conclusion}

In this work, we conduct a comprehensive study of directional source separation with the multi-channel microphone array on the geometry of smart glasses, demonstrating the effectiveness of incorporating directional information in source separation. In addition. Our experiments also imply that learning directional properties as a part of the neural network further improves voice quality. Lastly, we demonstrate that source separation benefits ASR performance, with joint training of source separation and ASR yields the lowest WER. One future work would investigate fairness and efficiency \cite{feng2022review} in source separation.